\documentclass[useAMS,usenatbib]{mn2e}

\usepackage{graphicx, color}
\newcommand{\apj}{ApJ}
\newcommand{\apjl}{ApJL}
\newcommand{\mnras}{MNRAS}
\newcommand{\aj}{AJ}
\newcommand{\nat}{Nature}
\newcommand{\araa}{ARA\&A}
\newcommand{\pasj}{PASJ}
\newcommand{\aap}{A\&A}

\title[Destruction of star clusters due to the radial migration in spiral galaxies]
{Destruction of star clusters due to the radial migration in spiral galaxies}

\author[M. S. Fujii and J. Baba]
{M. S. Fujii$^{1}$
\thanks{E-mail: fujii@strw.leidenuniv.nl(MSF);  baba.j.ab@m.titech.ac.jp(JB)} 
and J. Baba$^{2}$\footnotemark[1]\\
$^{1}$Leiden Observatory, Leiden University, NL-2300RA Leiden, The Netherlands\\
$^{2}$Interactive Research Center of Science, Tokyo Institute of Technology, 2-12-1 Ookayama, Meguro, Tokyo 152-8551, Japan}

\begin{document}

\date{Accepted 1988 December 15. Received 1988 December 14; in original form 1988 October 11}

\pagerange{\pageref{firstpage}--\pageref{lastpage}} \pubyear{2002}

\maketitle

\label{firstpage}

\begin{abstract}
Most stars in galactic disks are believed to be born as a member of star
clusters or associations. Star clusters formed in disks are disrupted
due to the tidal stripping and the evolution of star clusters
themselves, and as a results new stars are supplied to the galactic disks. 
We performed $N$-body simulations of star clusters in galactic 
disks, in which both star clusters and galactic disks are modeled as 
$N$-body (``live'') systems, and as a consequence the disks form transient 
and recurrent spiral arms.
In such non-steady spiral arms, star clusters migrate radially due
to the interaction with spiral arms.
We found that the migration timescale is a few hundreds Myr and that 
the angular momentum changes of star clusters are at most $\sim 50$\% 
in 1 Gyr.
Radial migration of star clusters to the inner region of galaxies
results in a fast disruption of the star clusters because of a 
stronger tidal field in the inner region of the galaxy. This effect is 
not negligible
for the disruption timescale of star clusters in galactic disks.
Stars stripped from clusters form tidal tails which 
spread over 1--2 kpc. While the spatial distribution of tidal tails 
change in a complicated way due to the non-steady spiral arms,
the velocity distribution conserve well even if the tidal tails are
located at a few kpc from their parent clusters. Tidal tails of
clusters in galactic disks might be detected using velocity plots.

\end{abstract}


\section{Introduction}

Star clusters are one of the fundamental building blocks of galactic 
disks because most stars are formed in star clusters 
\citep{2003ARA&A..41...57L}. In disk galaxies, new clusters 
born in galactic disks travel in their host disks experiencing 
disruptions and supply new stars to the disks. 
The disruption of clusters is caused by the tidal force from 
their host galaxy and also the internal evolution of star clusters 
themselves such as dynamical evolution, mass loss due to the 
stellar evolution, and gas expulsion \citep{2000MNRAS.318..753F, 
2003MNRAS.340..227B,2004AJ....127.2753D,2005AJ....129.1906C,
2005A&A...441..117L,2008MNRAS.384.1231B,2011MNRAS.413.2509G}. 

Non-axisymmetric structures in galactic disks, such as spiral arms 
\citep{2007MNRAS.376..809G} and bars \citep{2012MNRAS.419.3244B},
have also been expected to play important roles for the dynamical 
disruption of clusters. For example,
\citet{2007MNRAS.376..809G} investigated the effect of spiral-arm 
passages on the evolution of star clusters assumed to rotate in a 
fixed pattern speed, as in the stationary density wave theory 
\citep{1964ApJ...140..646L,1996ssgd.book.....B}.
However, self-consistent simulations of galactic disks 
have shown that self-excited spiral arms are not stationary regardless
of the existence of gas (or some kind of dissipation)
\citep{1984ApJ...282...61S,2002MNRAS.336..785S,2003MNRAS.344..358B,
2009ApJ...706..471B,2011ApJ...730..109F,2011MNRAS.410.1637S,
2011ApJ...735....1W,2012MNRAS.421.1529G,Baba+2012}. Such non-steady
spiral arms do not have a single pattern speed
but roughly follow the galactic rotation 
\citep{2011ApJ...735....1W,2012MNRAS.421.1529G}.
Therefore, these arms scatter stars everywhere in the disk by 
the co-rotation resonance 
\citep{2002MNRAS.336..785S,2012MNRAS.421.1529G,Baba+2012}.
Non-steady spiral arms change the gravitational fields around 
star clusters chaotically rather than periodically as 
in the stationary density waves.

In order to know the dynamical evolution of star clusters in 
galactic disks with non-steady spiral arms, we need to model both
star clusters and galactic disks as $N$-body (``live'') systems.
Such self-consistent $N$-body simulations are technically more 
difficult than those with rigid potential disks because 
the dynamical timescale of star 
clusters is much shorter than that of galactic disks and a large
number of particles are required for the modeling of the disks
\citep{2011ApJ...730..109F}. We solve this problem using
a direct-tree hybrid code, Bridge \citep{2007PASJ...59.1095F}.

In this letter, we perform self-consistent $N$-body simulations
of star clusters in live disks using Bridge and demonstrate that 
the angular-momentum exchange between star clusters and spiral arms 
causes the radial migration of star clusters of a few kpc from 
their initial galacto-centric radii. 
The migration timescale is shorter than the galactic rotation 
timescale, i.e., a few hundred Myr. 
The radial migration causes the tidal disruption of 
star clusters bringing them to closer to the galactic center.
Star clusters lose their mass in their perigalacticon passage,
and their tidal tails spread over a few kpc. 
We also find that tidal-tail stars stay close to their parent 
clusters in their velocity space even if
they are already a few kpc from the parent cluster. 
The tidal tail of young clusters might be detectable using 
their velocities.

\section{$N$-body simulations}

We performed a series of $N$-body simulations of star clusters 
embedded in a live galactic disk with spiral arms. 
We modeled both the disk and the clusters as $N$-body systems, 
but the halo is modeled as a potential. We set up the disks following models 
used in \citet{2011ApJ...730..109F}.
We adopted an exponential disk model with a total disk mass  
of $3.2 \times 10^{10} M_{\odot}$ with $3\times 10^6$ (3M) 
particles and as a consequence the mass of a disk-particle is 
$\sim 10^4 M_{\odot}$. The scale radius and scale height of the 
disk are 3.4 kpc and 0.34 kpc, respectively.
For the dark matter halo, we adopted the NFW model 
\citep{1997ApJ...490..493N} with the concentration
parameter of the halo, $c=10$. The virial radius and the mass of 
the halo are 122 kpc and $6.4\times 10^{11}M_{\odot}$.
We modeled star clusters as a King model with the dimensionless central 
potential $W_0=3$ \citep{1966AJ.....71...64K}. We adopted a total cluster 
mass of $10^5 M_{\odot}$ and a half-mass radius of 8 pc, and therefore the tidal 
radius is 30 pc. Our model is similar to young massive clusters in  
M51 and M82 (see Figure 9 in \citet{2010ARA&A..48..431P}) rather than 
those in in the Milky Way disk, which are more compact and therefore would 
be tidally disrupted less than our model. 
Included the stellar evolution, however, the Milky-Way clusters might 
expand a factor of 5--10 in the first 10 Myr as seen 
in observations (see Figure 8 in \citet{2010ARA&A..48..431P}). 
We used 8192 (8k) equal-mass particles for the cluster.

\begin{figure}
  \begin{center}
    \includegraphics[width=40mm]{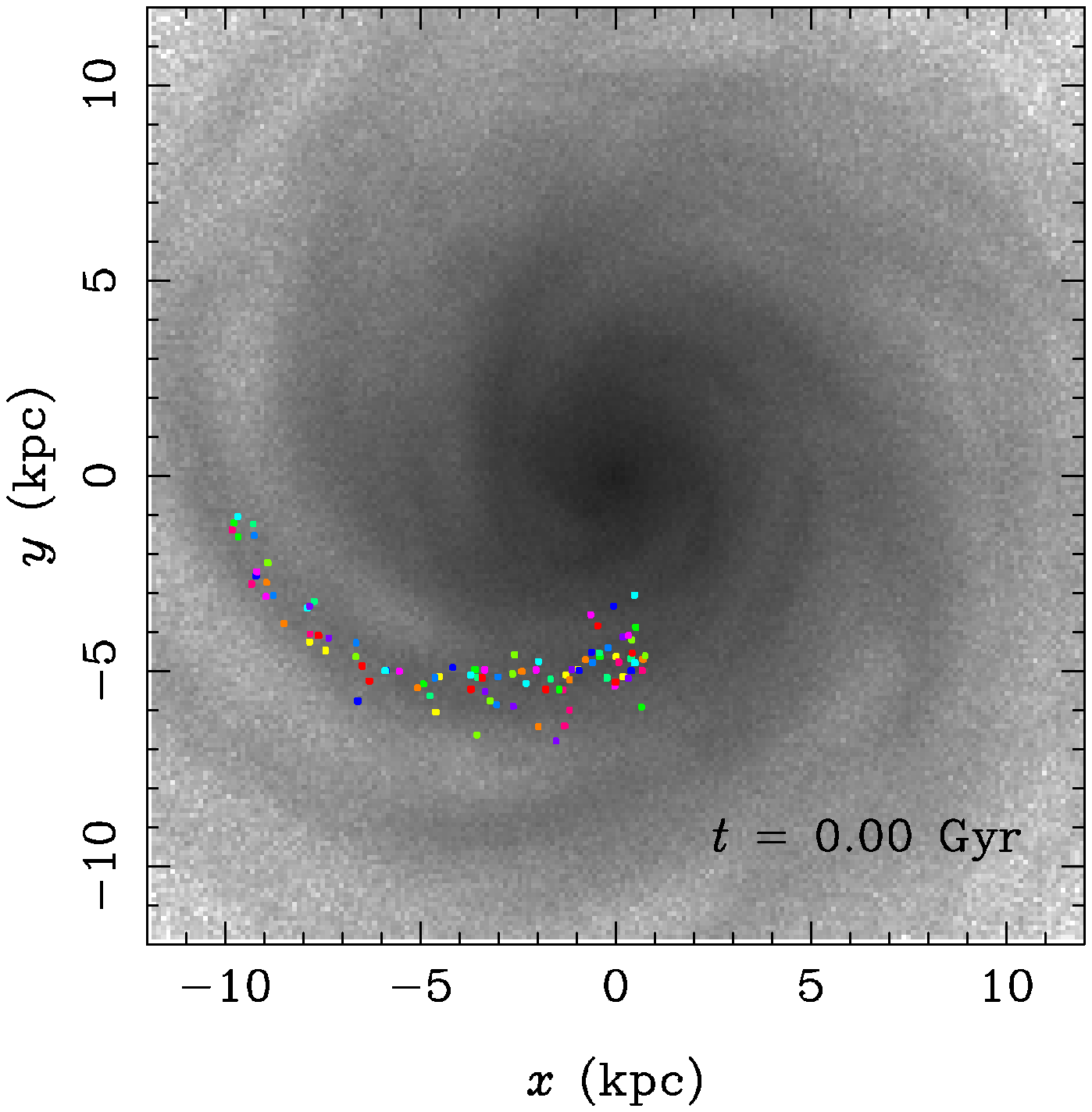}
    \includegraphics[width=40mm]{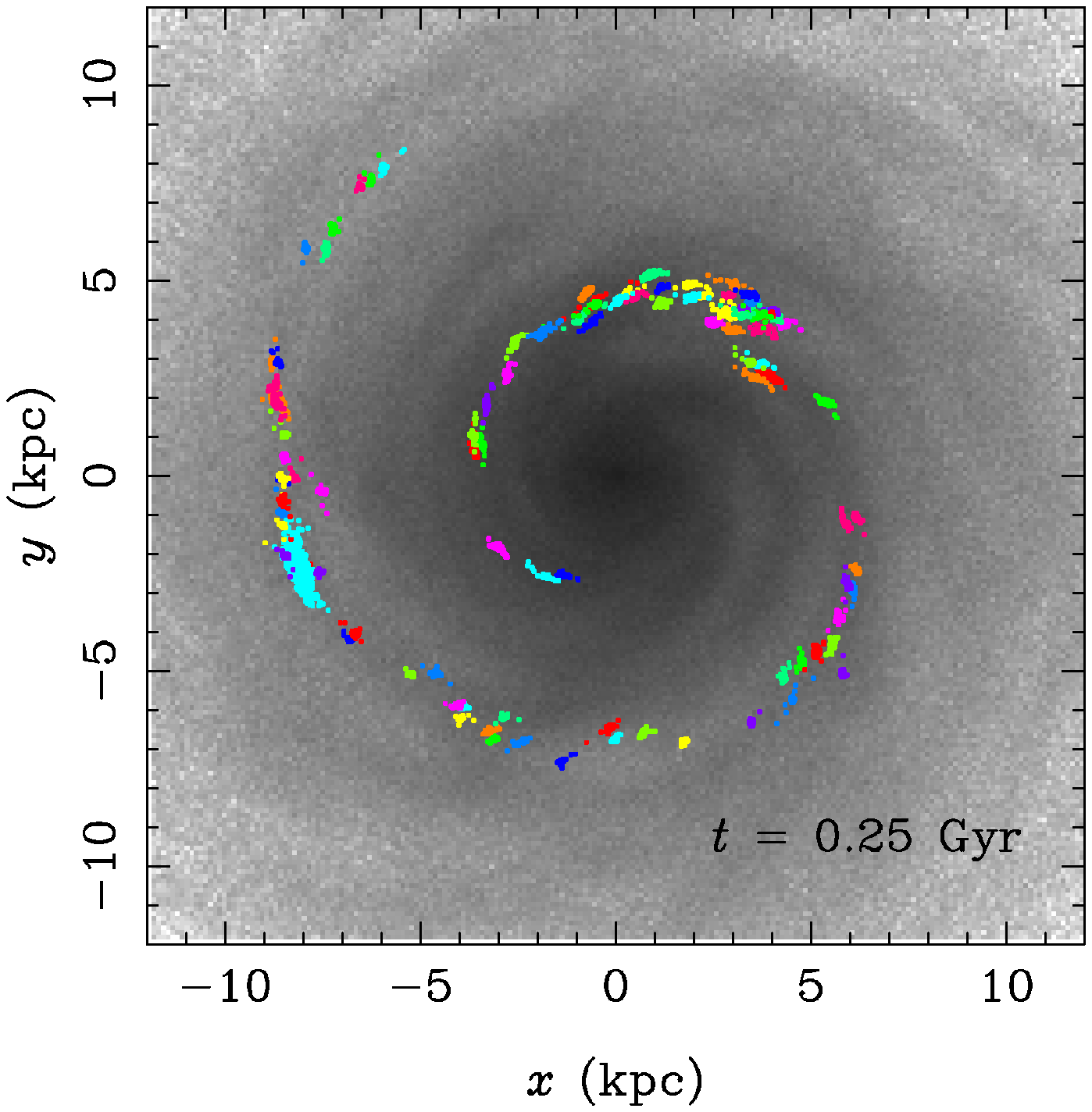}
    \includegraphics[width=40mm]{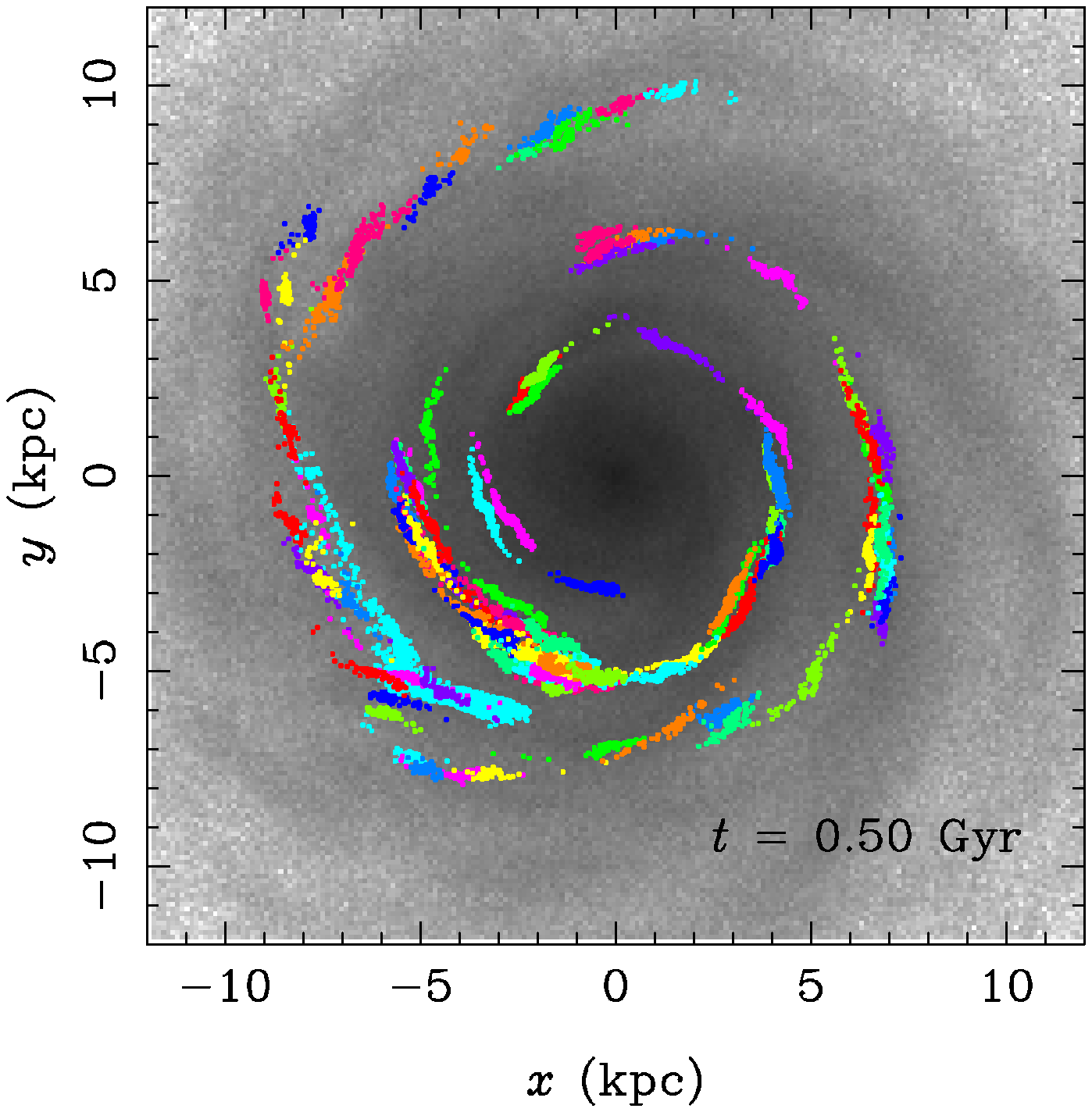}
    \includegraphics[width=40mm]{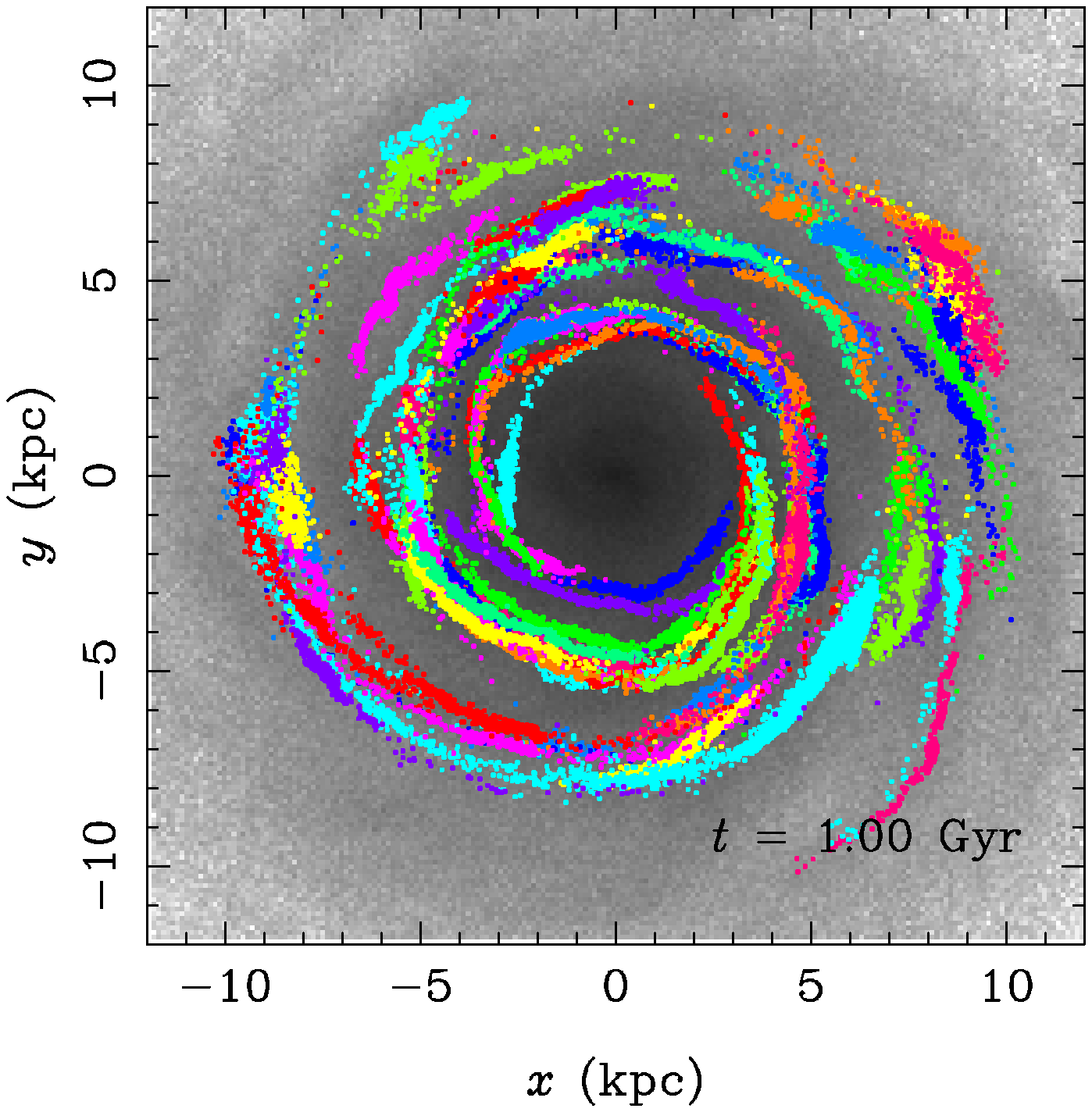}  
  \end{center}
  \caption{Snapshots of the initial condition and at 0.25, 0.5, and 1 Gyr. 
Gray scale shows the column density of the disk, and color points indicate 
cluster particles. Since these are only twelve colors but 97 clusters,
we plot several clusters with the same color. The 
color points are as large as the cluster size to make the points visible.
More than 80\% of cluster-particles are still bound at 1 Gyr. 
\label{fig:snap}}
\end{figure}

We first integrated only the disk up to 5 Gyr, in which self-excited 
spiral arms fully developed from the initial Poisson noise due to the
swing amplification. Then, we detected dense regions in the disk using 
the procedure below and put star clusters in the dense regions assuming 
that they are born there.
We detected disk particles whose eighth-nearest-neighbor position is 
closer than 70\% of the Jacobi radius of a cluster with $10^5 M_{\odot}$. 
We chose the densest position from the detected locations and rejected
other candidates within five Jacobi radii from the selected one 
in order to avoid that star clusters initially collide with each other.
Repeating this procedure, we chose 97 positions 
in a spiral arm in 4--10 kpc from the galactic center.
The initial positions of the star clusters are shown in the top left 
panel of Figure \ref{fig:snap}. 
We adopted the center-of-mass velocity of the eight 
neighbors as the cluster velocity.

The initial radial, azimuthal, and vertical velocity dispersion among 
the star clusters are 10.2, 7.7, and 7.5 $\rm km~s^{-1}$, respectively.  
Although these values are smaller than the mean of 
the disk stars (19, 14, and 14 $\rm km~s^{-1}$ at 8 kpc respectively),
star clusters would have rather smaller velocity dispersion than 
those of old stars if we assume that star clusters form from giant 
molecular clouds as is observed 
in the Milky Way \citep[the Geneva-Copenhagen Survey of the 
Solar Neighborhood;][]{2009A&A...501..941H}. 

The simulations are performed using 
a direct-tree hybrid code, Bridge \citep{2007PASJ...59.1095F}.
In Bridge, only the inner motion of star clusters are integrated 
using a direct $N$-body code and the other interactions are integrated 
using a tree code \citep{1986Natur.324..446B}.  
We adopted a sixth-order Hermite scheme for the direct method 
\citep{2008NewA...13..498N} without any softening and with an 
accuracy parameter of 0.9. For the disk particles, the gravitational 
potential is softened using Plummer softening with a
length of 10 pc. 
We adopted an opening angle of 0.4 with the center-of-mass 
approximation and a time step of 0.29 Myr for the tree code.

\section{Results}
\subsection{Radial migration of star clusters}

Figure \ref{fig:snap} shows the time evolution of the star clusters 
and disk. 
Although each spiral arm looks like a single coherent structure 
in each snapshot, it is transient and recurrent. 
They are wound up due to the differential rotation of the galactic disk 
and break up into multiple segments with a few kpc-scale, 
but the segments reconnect and form new long coherent arms 
(see online material movie). 

\begin{figure*}
  \begin{center}
    \includegraphics[width=130mm]{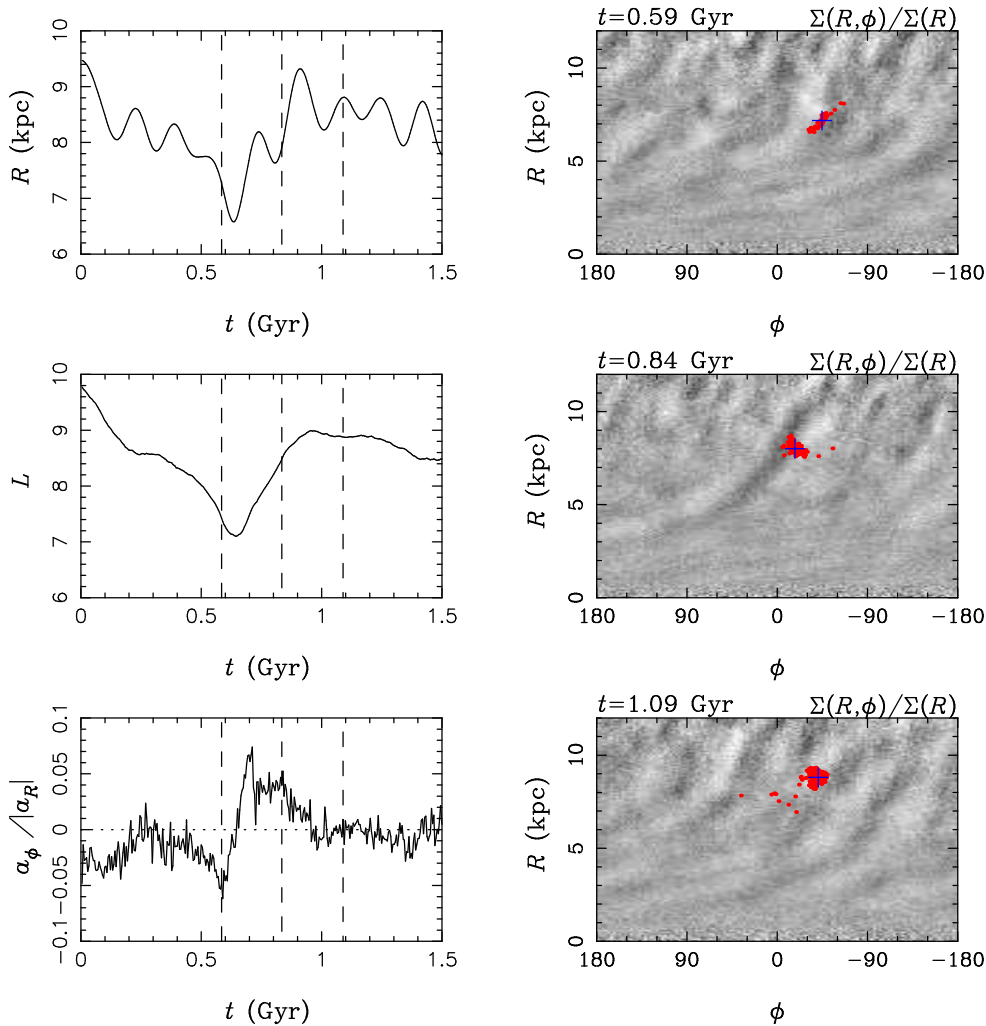}
  \end{center}
  \caption{
  {\it Left column}: Evolution of a star cluster. 
  Top: radial distance from the galactic center. 
  Middle: angular momentum. 
  Bottom: the force in azimuthal direction from the disk on the cluster 
  normalized by the force in radial direction. 
 {\it Right column}: position of the cluster particles (red points) and 
  the center of mass (blue cross) and the column density of spiral arms 
  in the $R-\phi$ plane at the times indicated by dashed lines in the left column. 
  From top to bottom the star cluster is located inter, 
  forward, and behind a spiral arm, respectively.   
  }
\label{fig:cluster76}
\end{figure*}

The orbits of star clusters in such transient spiral arms are 
complicated rather than a simple epicyclic motion.
The motion of star clusters in 
spiral arms is similar to that of gas and stars in the disk as shown 
in \citet{2011ApJ...735....1W,2012MNRAS.421.1529G}. Clusters tend 
to stay in or close to spiral arms and migrate along the arms. 
One of the orbits (the distance from the galactic 
center) obtained from our simulation is shown in the top left panel 
of Figure \ref{fig:cluster76} (see also online material movie). 
The cluster is initially located at around 9 kpc and migrates inward 
down to around 6.5 kpc losing its angular momentum, but it migrates
outward up again to 8.5 kpc (see the top left  
panel in Figure \ref{fig:cluster76}).

Such radial migration is caused by the dynamical interaction between star 
clusters and spiral arms. The bottom left panel of Figure 
\ref{fig:cluster76} shows time evolution of the azimuthal force 
from the disk on the star cluster. The cluster loses its 
angular momentum, when it is moving ahead of a spiral arm 
(see top right panel of Figure \ref{fig:cluster76}, 
in which spiral arms are moving from right to left). 
On the other hand, the star cluster gains angular momentum, 
when it is moving behind a spiral arm 
(see middle right panel of Figure \ref{fig:cluster76}). 
The angular momentum of clusters does not change when clusters are 
located at just the middle of two arms 
(see bottom right panel of figure \ref{fig:cluster76}). 
Thus, star clusters in non-steady disks lose or gain angular 
momentum and as a result migrate a few kpc from their initial 
positions. 

Such angular-momentum changes occur for all star clusters. 
The clusters in our simulation 
lost or gained at most $\sim 50$\% of their initial
angular momenta within 1 Gyr, which corresponds to radial migration 
of a few kpc.
The angular-momentum change in the clusters is similar to that of 
stars investigated in simulations of stellar disks 
\citep{2002MNRAS.336..785S,2008ApJ...675L..65R,2012MNRAS.421.1529G,
Baba+2012}.
Since star clusters in transient spiral arms migrate a few kpc  
in their galactic rotation timescale, i.e., a few hundred Myr,  
open clusters in the Galactic disk older than $\sim 100$ Myr 
are expected to have already migrated from their initial galactcentric 
radii. 

This rapid migration of star clusters may also change our 
understanding of the evolution of the Galactic disk.
Indeed open clusters in the Milky Way disk have been used as 
tracers for the dynamical and chemical evolution of the Galactic disk 
\citep[e.g.,]{1993A&A...267...75F,2003AJ....125.1397C,2009A&A...494...95M}.
In these studies, the spiral arms are considered to be stationary
density waves, but these results might change in non-steady arms.

\begin{figure}
  \begin{center}
    \includegraphics[width=80mm]{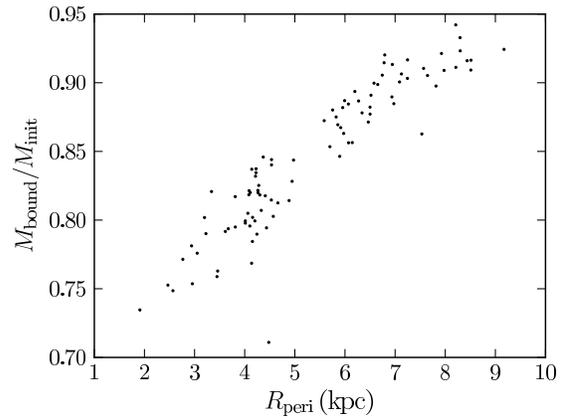}
  \end{center}
  \caption{
  The relation between the bound mass at 1.5 Gyr normalized by 
  the initial mass and the minimum perigalacticon distance. \label{fig:mass}}
\end{figure}

Another important effect of the radial migration is the tidal disruption
of star clusters due to the smaller Jacobi radii at smaller distance 
from the galactic center. In our models it changes from 64 pc at 8 
kpc to 45 pc at 4 kpc. In Figure \ref{fig:mass} we see a clear correlation
between the bound mass at the end of our simulation (1.5 Gyr) and 
the minimum perigalacticon distance of the clusters. In our simulation,
clusters lose masses mainly during their perigalacticon passages, 
and therefore the mass loss becomes larger when they migrates inward. 
However, internal heating due to the passage of spiral arms 
\citep{2007MNRAS.376..809G} does not work efficiently in the case 
of transient arms, because both star clusters and spiral arms are always 
corotating.

\subsection{Tidal tails}
Stars tidally stripped from their parent clusters form tidal tails
spreading over a few kpc (see Figure \ref{fig:snap}).
Their shapes are more complicated than those in simple 
spherical external potentials and change in time. 
When star clusters migrate from apogalacticon to 
perigalacticon, they move along a spiral arm, and their 
tidal tails also elongate along the spiral arm. 
When clusters are in the other phases (e.g., perigalacticon or 
apogalacticon), the tails come closer to the parent cluster again 
even if they are no longer bound to the cluster  
(see top right panel of Figure \ref{fig:cluster76}). 
This behavior is similar to that in the case in axisymmetric external 
potentials \citep{2005AJ....129.1906C}.

\begin{figure}
  \begin{center}
    \includegraphics[width=90mm]{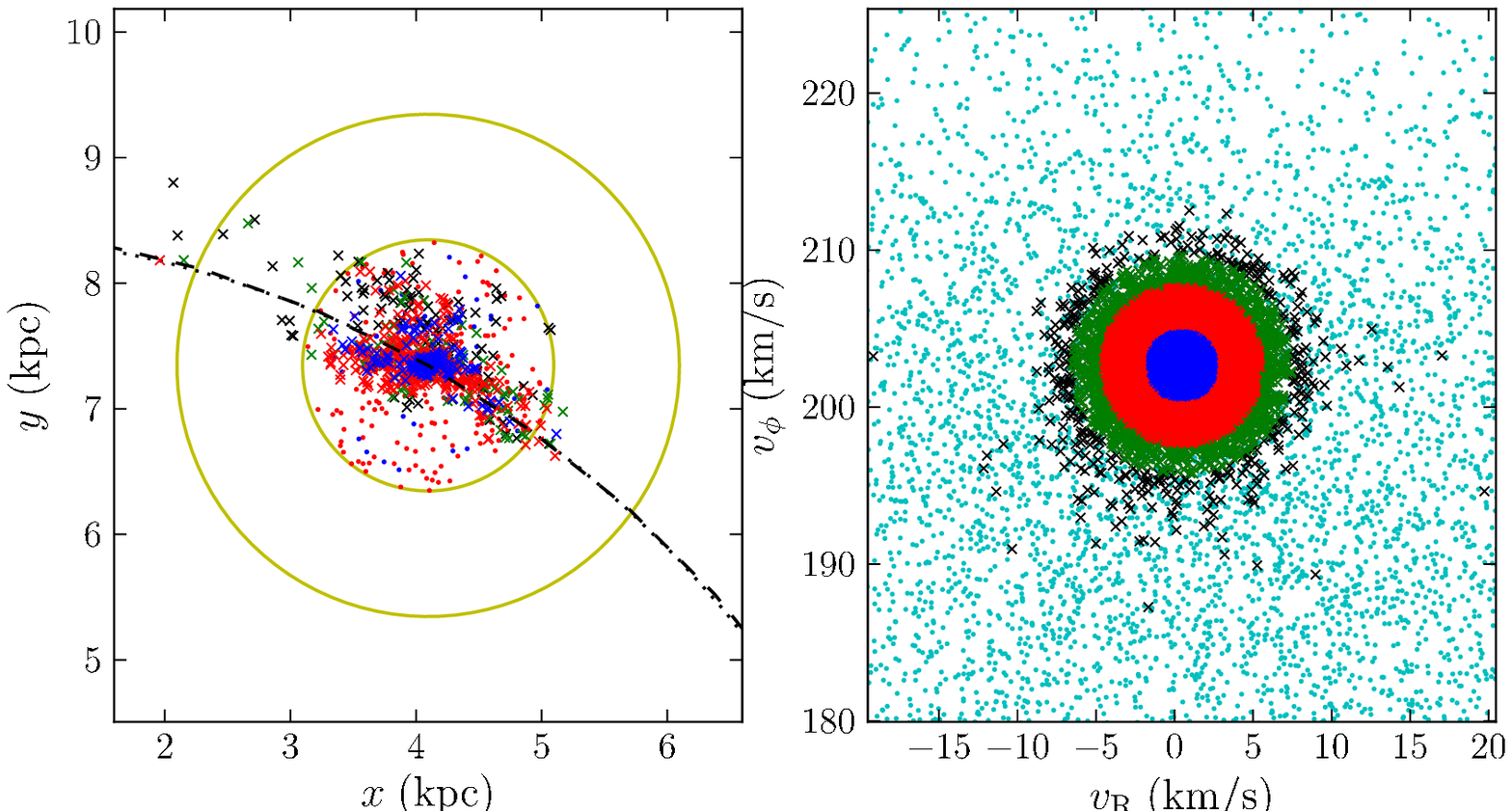}
    \includegraphics[width=90mm]{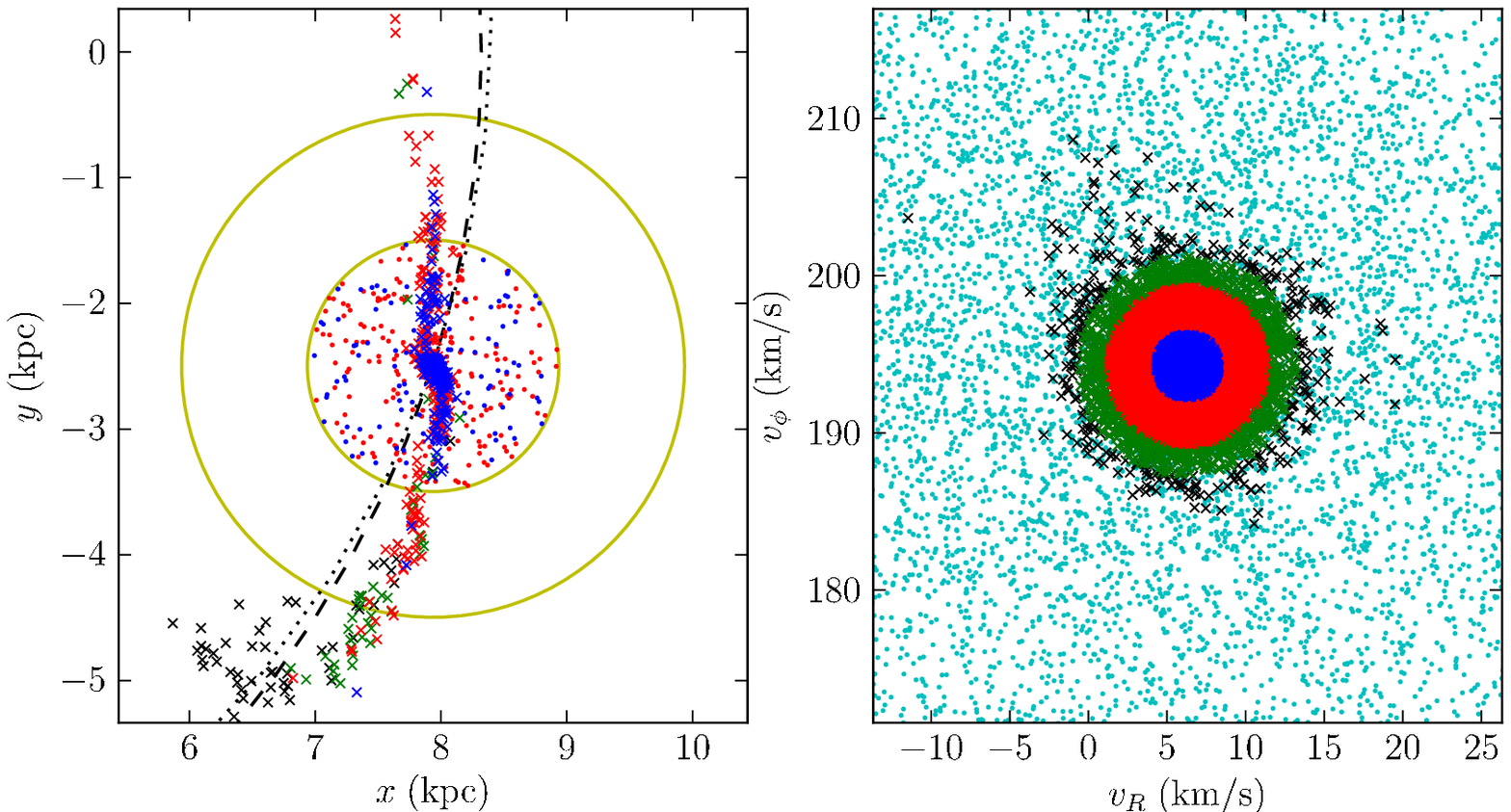}
  \end{center}
  \caption{Distribution of cluster particles and disk particles around 
clusters in $x$-$y$ plane (left) and $v_{R}$-$v_{\phi}$ phase (right). 
The cluster shown in the top panels is the same one shown in Figure 
\ref{fig:cluster76} at a time of 1.75 Gyr, when the cluster 
  is located at its perigalacticon. The bottom panels show a cluster, which is 
  moving from its perigalacticon to apogalacticon at a time of 1 Gyr. 
  Crosses and points indicate cluster and disk particles, respectively. 
  In the left panels, colors indicate the velocity shown in right panels. 
  Black dotted and dashed curves show orbits obtained from the simulation and 
  circular orbits at the distance of the cluster from the galactic center, 
respectively. 
  Yellow circles show the 1 and 2 kpc radius from the center-of-mass 
position of the cluster.  
  In the right panels, blue, red, green, and black crosses indicate cluster 
particles 
  with $<$2, 2--5, 5--7, and $>$7 $\rm km~s^{-1}$ from the mean velocity 
of the cluster particles. 
  Cyan points indicate disk particles which are located within 1 kpc from the 
  cluster center. Disk particles which are located within 1 kpc and have 
  velocities $<$2 and 2--5 $\rm km~s^{-1}$ from the mean velocity of 
the cluster are 
  shown in the left panels as blue and red points, respectively.\label{fig:xyv}}
\end{figure}

Figure \ref{fig:xyv} shows spacial (left) and velocity (right) 
distributions of two clusters. The cluster shown in the top panels is 
at its perigalacticon passage, and the one in the bottom panels is moving 
from its perigalacticon to apogalacticon.  
The cluster shown in the top panels is the same as that shown in 
Figure \ref{fig:cluster76}.
In spacial distribution plots (left panels in Figure \ref{fig:xyv}), 
cluster particles (crosses) show tidal 
tails spreading over a few kpc which are close to the orbits 
(black doted curves) or the circular orbits at the positions 
of the clusters (black dashed curve). In spite of such a large
spatial distribution, we find that the tidal-tail stars still remain 
close to their parent clusters in velocity space.
In the right panels, we plot the position of cluster stars 
in $v_{R}$-$v_{\phi}$ space as crosses. Colors in both
panels indicate the velocity deviation from the parent clusters; 
blue, red, green, and black indicate $<$2, 2--5, 5--7, and $>$7 
$\rm km~s^{-1}$, respectively. We also plot disk particles within 
1 kpc from the cluster center (cyan points). In contrast to the 
cluster particles, the disk particles have a larger distribution 
in the velocity spaces. 
In spatial plots we plot disk particle which are within 1 kpc from 
the cluster center and have velocity deviation of $<$5 $\rm km~s^{-1}$ 
in the velocity space (cyan points). 
If we detect star cluster tails using only their velocities, 
these stars would be detected as contamination. 
They are located randomly in spatial plots, while star cluster 
particles distribute in tidal tails. 

\section{Summary}

We performed a series of $N$-body simulations of star clusters in live 
stellar disks with multiple spiral arms. In these simulations, both 
galactic disks and star clusters are modeled as $N$-body systems and 
integrated self-consistently.
Our results show that star clusters migrate radially a few kpc in 
the time scale of their orbital period in the disk (a few 100 Myr)
because of the angular-momentum exchange with transient spiral arms.
The angular-momentum change of the clusters is at most $\sim 50$\% 
of the initial angular momentum within 1 Gyr. 

In the case of non-steady transient spiral arms, the radial migration strongly 
affects the tidal disruption of star clusters because star clusters 
lose more mass when they approach the galactic center due to the 
smaller Jacobi radii. This kind of disruption mechanism does not 
appear in stationary density waves. 
The heating due to the non-steady (corotating) spiral-arm passage 
would not be as strong as that by the density-wave spiral arms, 
because the adiabatic change of the energy due to the slow passages 
of spiral arms suppresses the heating per passage, and  the number 
of spiral passages is quite few (see the corotation case in Figure 8 
in \citet{2007MNRAS.376..809G}).
Furthermore, the radial migration of star clusters can
carry stars far from their original orbital radii and finally
the distribution of stars would be much wider than those in the case 
of the density waves \citep{2010ApJ...713..166B}.

With transient spiral arms, star clusters and their tidal tails tend
to stay in or close to spiral arms. The shape of tidal tails of 
clusters change in a complicated way in time compared 
to those in a smooth tidal field like a halo potential.
When a star cluster approach the galactic center, star clusters 
move along a spiral arm as is the case of stars and gas moving in 
spiral arms. 
In this phase, the tidal tails also spread along the spiral arm. 
During the apogalacticon passage, on the other hand, the tidal tails 
are compressed and distribute around 1 kpc from the cluster even 
though they are unbound. 
The tidal tails of clusters might be detectable even if they 
spread over a few kpc, because the tidal-tail stars
still remain very close to the cluster in velocity space after they
become unbound. 
If we know the velocity of stars, we might be able to detect the 
tidal tails of star clusters in the Galactic disk using 
a future astrometry such as {\it Gaia} and {\it JASMINE}.

\section*{Acknowledgments}
The authors thank Dan Caputo and Jeroen B\'{e}dorf for careful 
reading of the manuscript. 
This work was supported by Postdoctoral Fellowship for Research Abroad 
of the Japan Society for the Promotion of Science (JSPS) and HPCI
Strategic Program Field 5 ``The Origin of Matter and the Universe.'' 
Numerical computations were carried out on GRAPE-DR at the
Center for Computational Astrophysics (CfCA) of the National 
Astronomical Observatory of Japan and the Little Green Machine 
at Leiden University.

\bsp

\label{lastpage}

\end{document}